\documentclass[12pt,preprint]{emulateapj}
\usepackage{natbib}
\usepackage{graphicx}
\usepackage{amssymb}
\usepackage{epstopdf}
\usepackage{subfigure}

\begin{document}

\shorttitle{Cosmic Ray Anisotropy and Heliospheric Boundaries}
\shortauthors{Desiati \& Lazarian}

\title{Anisotropy of TeV Cosmic Rays and the Outer Heliospheric Boundaries}
\author{P. Desiati}
\affil{Wisconsin IceCube Particle Astrophysics Center (WIPAC)\\Department of Astronomy, University of Wisconsin, Madison, WI 53706 }
\author{A. Lazarian}
\affil{Department of Astronomy, University of Wisconsin, Madison, WI 53706 }

\begin{abstract}
Cosmic rays in the energy range from about 10's GeV to several 100's TeV are observed on Earth with an energy-dependent anisotropy of order 0.01-0.1\%, %$10^{-4}-10^{-3}$,
and a consistent topology that appears to significantly change at higher energy. The nearest and most recent galactic cosmic ray sources might stochastically dominate the observation and possibly explain a change in orientation of the anisotropy as a function of energy. However, the diffusion approximation is not able to explain its non-dipolar structure and, in particular, the significant contribution of small angular scale features. Particle propagation within the mean free path in the local interstellar medium might have a major role in determining the properties of galactic cosmic rays, such as their arrival distribution. In particular, scattering on perturbations induced in the local interstellar magnetic field by the heliosphere wake, may cause a re-distribution of anisotropic cosmic rays below about 100 TeV toward the direction of the elongated heliotail and of the local interstellar magnetic field in the outer heliosphere. Such scattering processes are considered responsible of the observed TeV cosmic ray global anisotropy and its fine angular structure.
\end{abstract}

%-----------------------------------------------------------------------------------------------------------------------------------------------
\section{Introduction}
\label{sec:intro}

Galactic cosmic rays possess a small but significant anisotropy of order $10^{-4}-10^{-3}$. Observations in the northern hemisphere were reported from energies of tens to several hundreds GeV with muon detectors~\citep{nagashima,hall99,munakata11}, in the 1-10's TeV energy range with Tibet AS$\gamma$ array~\citep{amenomori,amenomori11}, Super-Kamiokande \citep{guillian}, Milagro \citep{abdo} and ARGO-YBJ~\citep{argo,shuwang11}, up to 100's TeV by EAS-TOP~\citep{aglietta}. The IceCube Observatory recently reported similar observations in the southern hemisphere in the 10-100 TeV energy range~\citep{abbasi,abbasi11}, also using the IceTop surface array component~\citep{santander12}. The first major feature of the cosmic ray anisotropy is that it is more complex than a dipole, and that its amplitude seemingly increases with energy up to the 10's TeV range, above which it decreases again~\citep{amenomori05}. The second interesting characteristics is a significant change in topology at energies in excess of approximately 100 TeV (as observed by EAS-TOP and IceCube). And the third major property of the observations is the discovery of significantly small angular features in the TeV energy range.

The origin of such uneven global distribution in the arrival direction of galactic cosmic rays is still unknown. A dipole anisotropy would be expected from the motion of the solar system with respect to the cosmic ray plasma rest frame. If cosmic rays do not co-rotate with the Galaxy, an observer on Earth would see an apparent excess of particle counts toward the direction of galactic rotation and a deficit in the opposite direction. Such a convective anisotropy is referred to as the Compton-Getting effect~\citep{compton, compton2}. The predicted amplitude is expected to be of order 10$^{-3}$ with the apparent relative excess in right ascension at about 21 hr. Apart from the fact that the observed anisotropy is not a simple dipole, neither the expected amplitude nor the phase seem to be consistent with the observations. On the other hand, the reference frame of the galactic cosmic rays is not known, therefore it is reasonable to assume that the Compton-Getting effect could be one of the contributions to the cosmic ray anisotropy.

It has been speculated that the observed anisotropy might be a natural consequence of the stochastic nature of cosmic ray galactic sources, in particular nearby and recent Supernova Remnants (SNR). Discreteness of such sources, along with cosmic ray propagation through a turbulent interstellar medium, might lead to significant fluctuations of their intensity in space and, therefore, to a faint broad anisotropy (see for instance~\cite{erlykin,blasi,pohl12}). In absence of signatures of nearby sources of cosmic rays an anisotropy could be generated by convection from persistent cumulative magnetized flow field of old SNRs~\citep{biermann12}. Or it can result from the combined effects of the regular and the turbulent (fluctuating) magnetic fields in our vicinity~\citep{battaner}. However, the solar system is surrounded by a highly heterogeneous Local Interstellar Medium (LISM)~\citep{frisch2011a}, that can affect the arrival direction distribution of the cosmic rays. The common velocity of the nearby interstellar clouds could indicate that they are part of an evolved sub-shell of the superbubble associated to the Loop I, expanding from the Scorpion-Centaurus Association. Various indirect determinations of the Local Interstellar Magnetic Field (LIMF) suggest that its direction is coherent over scales of about 100 pc and roughly parallel to the local surface of Loop I shell~\citep{frisch2010b,frisch2011b,frisch2012,frisch2012b}. The observations, therefore, could be associated to cosmic ray diffusive streaming  between colliding nearby interstellar clouds of the expanding Loop I shell~\citep{schwadron12}. Since the solar system is located almost at the edge of the so-called Local Interstellar Cloud (LIC), a partially ionized cloudlet within the Local Bubble, it was proposed by~\cite{amenomori2,amenomori11b} that a non-dipolar anisotropy could be generated by the diffusion of cosmic rays through the LIMF connecting the solar system to the interstellar medium outside the LIC. 
On the other hand, the observation of a topological change of the anisotropy pattern at an energy in excess of about 100 TeV, where a relative deficit is observed in the region of the sky dominated by a broad excess at lower energies, is an indication of a phenomenological transition in the cause of high energy cosmic rays anisotropy.
%Whatever causes the anisotropy at lower energy, it has no effect above about 100 TeV.
A proton in a 3 $\mu$G magnetic field has a maximum gyro-radius $R_g\approx 80\cdot E_{TeV}$ AU\footnote{an Astronomical Unit (AU) is the mean distance between Earth and the Sun, corresponding to approximately 1.5$\times$10$^{11}$ m.}. Even though the mean free path of cosmic rays in magnetized plasma can be significantly larger~\citep{yan08}, it is reasonable to believe that cosmic rays with energy below about 100 TeV must be affected by the heliosphere, considering its extension from hundreds to several thousands AU~\citep{pogorelov,pogorelovb,izmodenov03,izmodenov}.

The anisotropy of TeV cosmic rays shows evidence of statistically significant small angular features. By effectively averaging the cosmic ray counts over an angular range of a few tens degrees, and subtracting the averaged map from the observation, the regions where cosmic ray intensity changes significantly within such angular scale are identified as localized fractional excess or deficit regions. Such technique led to the discovery of two highly significant localized fractional excess regions in the northern hemisphere by Milagro~\citep{abdo2}, also observed by Tibet AS$\gamma$~\citep{amenomori} and ARGO-YBJ~\citep{vernetto, iuppa11}. Similar observations were recently reported in the southern hemisphere by the IceCube Observatory as well~\citep{abbasi11b}. Although astrophysical arguments have been raised to explain the observed small scale features~\citep{salvati,drury,salvati2}, also in connection to cosmic ray propagation through the turbulent magnetic field~\citep{malkov}, the existence of such localized regions in the arrival direction distribution of TeV cosmic rays is a phenomenological indication that processes occurring within the mean free path are likely responsible for the observations. For instance, scattering of TeV-PeV cosmic ray particles with the cascading turbulent LIMF within a few 10's pc can generate intermediate and small scale perturbations over an underlying global anisotropy~\citep{giacinti}. However the observed small scale anisotropy features at 1-10 TeV energy range, do not seem to have a stochastic nature, but they rather appear to be correlated with the global anisotropy structure.

The change in topology of the cosmic ray anisotropy above about 100 TeV, is observed at the energy scale where the influence of the heliosphere is expected to be sub-dominant. Assuming that the anisotropy of TeV cosmic rays approaching the heliosphere is the same as that above 100 TeV,
%TeV cosmic ray anisotropy is the same they have at high energy while approaching the heliosphere,
this suggests that the arrival distribution is globally re-shaped by heliospheric perturbations induced in the LIMF, and it might explain some apparently coincidental correlation between the anisotropy and the outer heliospheric features. In this paper, for the first time, combined full-sky observations of the global anisotropy of 1-10 TeV cosmic rays are used to provide some evidence that the change in topology at about 100 TeV is due to the heliosphere, which is also responsible for the small scale angular structure. In Sec.~\ref{sec:helio} the heliospheric magnetic field structure is illustrated. In Sec.~\ref{sec:obs} a more detailed description of the experimental observations is provided and Sec.~\ref{sec:probe} reports the discussion of the proposed model.

%-----------------------------------------------------------------------------------------------------------------------------------------------
\section{The Heliosphere}
\label{sec:helio}

The solar system's motion through the LIC produces a complex interface due to the interaction between the supersonic solar wind and the interstellar flow, called the heliosphere. A termination shock, where the solar wind pressure equals that from the interstellar flow, is formed at approximately 100 AU from the Sun. The interface separating interplanetary and interstellar magnetic fields, called heliopause, is at a distance of approximately 200 AU in the upstream direction (the heliospheric nose) and it has a comet-like shape downstream due to the heliospheric plasma streamlines deviated by the interstellar flow (the heliospheric tail, or heliotail). The heliotail can be as wide as about 600 AU and as long as several thousands AU~\citep{izmodenov,pogorelov}. Deformations on both the heliosphere and the LIMF draping around it are produced by their mutual interaction~\citep{pogorelovi,pogorelovii,pogorelovb}. The heliospheric magnetic field has been studied with detailed Magneto-Hydrodynamic (MHD) simulations, where the effects from the 26 day solar rotation and the 11 year solar cycle were considered (see~\cite{pogorelov}). %The heliospheric plasma flow deviates magnetic field lines toward the tail, where the subsonic solar wind speed is of order 100 km/sec and decreasing until the tail dissolves into the interstellar medium.
Over the solar cycle the magnetic field polarity is reversed every 11 years, generating unipolar regions dragged along the heliotail by the $\sim$ 100 km/s solar wind~\citep{par79}. In particular these regions grow to their maximum latitudinal extent during solar minimum (about 200-300 AU in size) and reduce to zero at solar maximum, during which the heliospheric plasma is dominated by the strongly mixed polarity domains (about 0.1-1 AU in size) from solar rotation~\citep{nerney}. Due to the tilt of the solar magnetic axis with respect to its rotation axis, the unipolar regions are thinner at lower latitudes. Therefore the tailward line of view is dominated by the finely alternating magnetic field, while along sightlines away from it the magnetic domains have larger size\footnote{the actual solar magnetic field structure and dynamics is significantly more complex~\citep{luhmann11}, thus affecting the detailed unipolar domains conformation, although the general properties can still be described in terms of a tilted dipolar magnetic field of the Sun.}. MHD numerical simulations show that the sectored unipolar magnetic field regions can propagate along the tail for several solar cycles before they dissipate into the LISM. The periodic variations induced by solar cycles on the termination shock and the overall heliospheric plasma have profound effects on the dynamic behavior of the heliosphere~\citep{zank03,washimi11,washimi12}. The corresponding magnitude of the LIMF can vary about 25\% and the Alfv\'en velocity by about 20\%~\citep{pogorelov}.

There is observational evidence that the plasma in the heliosheath has Reynolds number $Re \approx$ 10$^{14}$ (see~\cite{lazopher} and references therein), meaning that the strength of non-linear convective processes at the largest scale is more important than the damping viscous processes in the dynamics of the flow. We expect a similarly high Reynolds number in the inner heliotail as well, and we know that this is true for the interstellar medium in general. In such conditions it is very unlikely that plasma flow stays laminar. In particular the wake downstream the interstellar flow develops turbulence at an injection scale of the order of the heliotail thickness (i.e. $\approx$ 600 AU). In addition, the presence of neutral atoms in the partially ionized LISM is essential for the dynamics of the heliosphere and LIMF interaction. Charge-exchange processes between the interstellar inflowing neutral atoms and the outflowing solar wind protons can produce Rayleigh-Taylor type instabilities on the heliopause with amplitude of a few tens AU and over a time scale of a few hundreds years~\citep{liewer96,zank96}. In a model of plasma-neutral fluid coupled via collision and charge-exchange processes, it is found that such non-linear coupling leads to alternate growing and damping of Alfv\'enic, fast and slow turbulence modes, at $L \sim$100's AU scale and with evolution time longer than inertial time $L/V_A$, with $V_A$ the Alfv\'en velocity~\citep{zank99,florinski05,borovikov08,zank09,zank}.

The LIMF is deformed by the heliosphere over the largest scale of about 600 AU (i.e. the heliosphere transversal size), which is much smaller than the injection scale of the interstellar turbulence. The same non-linear processes that produce turbulence on the wake are responsible for its cascading down to smaller scales. Since turbulence evolves along hydrodynamic Kolmogorov cascades, the Alfv\'en velocity $\delta V_l$ of heliospheric magnetic perturbations at scale $l$ can be expressed as
\begin{equation}
\delta V_l = V_h\,\left(\frac{l}{L_h}\right)^{1/3},
\label{eq:alfven}
\end{equation}
where $V_h\sim$ 23-26 km/sec is the interstellar flow velocity (see~\cite{ibex12} and references therein) and $L_h\approx$ 600 AU the injection scale of the heliospheric induced turbulence. Since the Alfv\'en velocity in the partially ionized plasma within the LIC that surrounds the heliosphere is $V_A\sim$ 13-17 km/sec~\citep{spangler}, turbulence is super-Alfv\'enic at scales $l_A \gtrsim L_h (V_A/V_h)^3 \approx$ 80-250 AU. Magnetic perturbations at such scales are strong (i.e. $\delta B/B_0 \approx 1$) and cannot propagate into the upstream interstellar flow direction (i.e. they trail downstream). Super-Alfv\'enic turbulence isotropically cascades down to the scale $l_A$, where it becomes sub-Alfv\'enic and the mean magnetic field gets dynamically important.
Cascading sub-Alfv\'enic turbulence is critically balanced, making perturbations more elongated along the magnetic field at small spatial scales, while keeping the energy cascade rate constant~\citep{GS95}. This is true for incompressible turbulence, however it can be considered valid in the compressible case as well (see~\cite{LB06}), which is what is expected to occur in realistic astrophysical plasmas. In this case turbulence can be composed by magnetosonic modes, in addition to shear Alfv\'en modes. On the other hand, turbulent dissipation in the cascading process has the effect of suppressing magnetosonic waves and effectively makes turbulence nearly incompressible~\citep{zank06}. As a consequence, compressible magnetosonic modes decay quickly leaving only the shear sub-Alfv\'enic modes to dominate the energy cascades.

In the partially ionized very local interstellar medium, ion-neutral collisions is a source of damping mechanism for Alfv\'enic turbulence, and the collisional mean free path in the LISM is approximately 300 AU~\citep{spangler}. Therefore we can picture the outer heliospheric boundary as populated by cascading turbulent ripples in the LIMF trailing downstream (at the largest scales) and with exponentially decreasing amplitude as a function of distance outward into the interstellar medium.

%-----------------------------------------------------------------------------------------------------------------------------------------------
\section{Observations of Cosmic Ray Anisotropy}
\label{sec:obs}

\begin{figure}[!t]
\begin{center}
\includegraphics[width=\columnwidth]{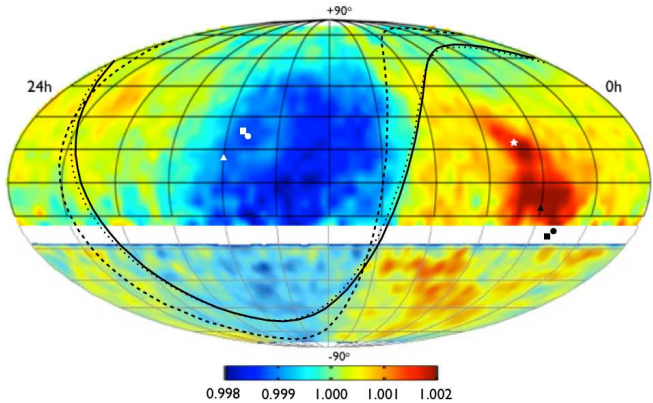}
\caption{\small Map in equatorial coordinates constructed by combining relative intensity distributions of cosmic ray counts independently normalized within declination bands of order $1^{\circ}-5^{\circ}$. The map shows the observation by Tibet AS$\gamma$ at about 5 TeV in the northern hemisphere~(from \cite{amenomori11}) and by IceCube at a median energy of 20 TeV in the southern hemisphere~(from \cite{abbasi}). Note the different characteristic energy of the cosmic ray particles in the two hemispheres. Also shown the equator "at infinity" (see text) in black continuous line, corresponding to the direction of LIMF inferred in~\cite{funsten} (circular symbols), that in black dotted line from~\cite{schwadron} (square symbols), and that in black dashed line from~\cite{heerikhuisen} (triangle symbols). The star symbol indicates the downwind interstellar medium flow with coordinate (5 hr, +17$^{\circ}$). Color scale indicates relative excess (red) and deficit (blue) with respect to average intensity, and the map grid is 2 hr in right ascension and 15$^{\circ}$ in declination.}
\label{fig:LargeScale2}
\end{center}
\end{figure}

%The observations of sidereal anisotropy of cosmic rays from tens of GeV to the multi-TeV energy range in the northern equatorial hemisphere, show a broad excess toward the direction that includes the heliotail (the so-called tail-in excess), while the opposite side of the northern equatorial sky is characterized by a region of deep deficit (the so-called loss cone) and a region of relative mild excess, at different azimuthal ranges with respect to the LIMF. Some experiments that measure the arrival direction of the cosmic rays via the extensive air showers they induce in the atmosphere, have some contamination from events induced by TeV gamma rays. In fact the mild excess region in the northern equatorial sky also includes a relatively localized region generated by gamma rays from Cygnus (Tibet, milagro).

%The only observation of sidereal anisotropy in the southern equatorial hemisphere in the multi-TeV energy region was reported by the IceCube Observatory (reference). This result shows a smooth continuation of the tail-in excess and the loss-cone deficit across the southern hemisphere, along with an evidence of a milder relative excess region toward the nose of the heliosphere.

The observed global cosmic ray anisotropy from tens of GeV to tens of TeV energy range, is generally characterized by a broad relative excess between right ascension $\alpha \approx$ 18 hr and $\alpha \approx$ 8 hr, and a relative deficit in the rest of the sky, as shown in Fig.~\ref{fig:LargeScale2}. The figure shows the map in equatorial coordinates, of cosmic ray relative intensity in arrival direction observed by Tibet AS$\gamma$ at median energy of 5 TeV in the northern hemisphere~\citep{amenomori11}, and by IceCube at about 20 TeV median energy in the southern hemisphere~\citep{abbasi}. The relative intensity measures the deviation of cosmic ray counts in a given declination band with respect to the average count in that band. Relative intensity distributions are independently determined for each declination band (of order $1^{\circ}-5^{\circ}$) and are used to construct the map, thus providing information on the relative modulation of the arrival direction of cosmic rays along the right ascension. Although the right ascension position of the relative excess and deficit regions as a function of declination depends on the cosmic rays energy, the global pattern of the anisotropy does not change significantly across the sky up to $\approx$ 100 TeV. Moreover the energy response distribution for these observations are reported to be rather wide, therefore the different characteristic energies corresponding to the two halves of the sky are partially mitigated by a significant overlap in energy response. %Consequently the combined sky map in Fig.~\ref{fig:LargeScale2} can provide a valuable tool to study the 1-10 TeV cosmic ray anisotropy.

The relative excess region in the northern hemisphere shows a complex angular structure around the apparent direction of the interstellar downstream flow direction (approximately the inferred direction of the heliospheric tail), with equatorial coordinates ($\alpha$, $\delta$) $\approx$ (5 hr, +17$^{\circ}$) (indicated with a white star in the figure). This is the region where a broad enhancement denominated tail-in excess was observed for the first time with sub-TeV cosmic rays~\citep{nagashima}. Although the global relative excess observed in the southern hemisphere seems to have a different fine structure, it appears topologically connected to that observed in the north. This might be an indication that even though the energy distributions of the two observations overlap, some small scale differences could still arise, perhaps also related to different mass composition sensitivity of the two experiments. The visible feature at equatorial coordinates ($\alpha$, $\delta$) $\approx$ (21 hr, +32$^{\circ}$) is reported to be originated by high energy gamma rays from the Cygnus region to which the Tibet AS$\gamma$ is also sensitive.

%\begin{figure}[h!]
%\begin{center}
 %\subfigure[] % caption for subfigure a
%{
 %     \includegraphics[width=\columnwidth]{figs/Tibet_12_TeV.eps}
%}
%\hspace{1cm}
% \subfigure[] % caption for subfigure a
%{
 %     \includegraphics[width=\columnwidth]{figs/IceCube-22_20_TeV.eps}
%}
%\hspace{1cm}
%
%\caption{
%\label{fig:LargeScale}
%Maps of the relative intensity of the cosmic ray arrival distribution as observed by the Tibet Air Shower at about 12 TeV~(extracted from \cite{amenomori}) (a) and by the IceCube Observatory in its 22 string configuration at a median energy of 20 TeV~(from \cite{abbasi}) (b). The color scale is the relative intensity value for each bin in the map normalized to unity within each declination band. A gaussian smoothing with 5$^{\circ}$ width was used for the IceCube map.}
%\end{center}
%\end{figure}

\begin{figure}[t]
\begin{center}
\includegraphics[width=\columnwidth]{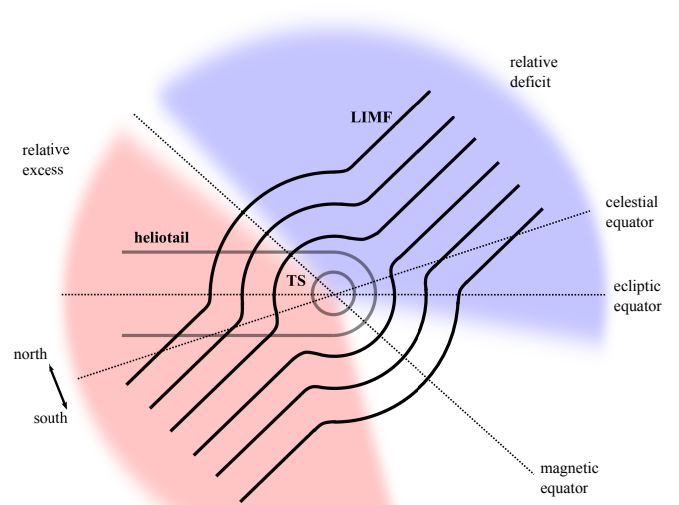}
\caption{\small Schematic representation of the elongated heliosphere with respect to the LIMF draping around it. The arrival distribution of cosmic rays with energy below 100 TeV have a relative excess in the portion of the sky south of the magnetic equator that includes the heliospheric tail in the downstream interstellar flow direction (represented in red). The deficit is observed toward the upstream interstellar flow direction, north of the magnetic equator (represented in blue). The regions with the largest gradient in the cosmic ray intensity (in white between the excess and deficit) are approximately located along the magnetic equator "at infinity" (see text). Also shown the celestial and ecliptic equators.
%Schematic representation of the relative location of the heliosphere and LIMF draping around the heliosphere, and the approximate position of the observed broad excess (in red) and deficit (in blue) regions of the multi-TeV cosmic ray arrival direction distribution. The white sectors indicate the transition between the two regions: such transition approximately located by the magnetic equator and it is steeper in regions with angular proximity to the heliotail (see text). In the figure the projected equatorial and ecliptic equators are highlighted, along with the magnetic equator defined as the locus of points that is equi-distant from the magnetic poles.
}
\label{fig:cartoon}
\end{center}
\end{figure}

In general, the combined observations reported in Fig.~\ref{fig:LargeScale2} can provide a valuable tool to study the global 1-10 TeV cosmic ray anisotropy, although consistent observations with comparable energy and mass sensitivities are desirable. It is interesting to note that such anisotropy structure is not compatible with a simple dipole%(see~\cite{amenomori2, amenomori11})
, therefore the cause cannot be simply from diffusive propagation across the Galaxy. The figure shows the orientations of the LIMF as inferred by various interpretative models of the keV Energetic Neutral Atoms (ENAs) observations with IBEX~\citep{funsten,schwadron,heerikhuisen} along with their corresponding equator "at infinity"\footnote{equator "at infinity" is the plane perpendicular to the magnetic field lines and passing by the observer's location, assuming the magnetic field is uniform.}. These orientations are consistent with that based on Ca II absorption lines and on observations of H I lines in white dwarf stars~\citep{frisch1996}, with polarization data~\citep{frisch2012,frisch2012b}, and on observation of radio emissions from the inner heliosheath and from observations of hydrogen deflection~\citep{lallement, opher2007, lallement10}. If we assume that the various indirect probes provide an accuracy of order 30$^{\circ}$ on the LIMF direction, we note that the magnetic equator is approximately located along a big portion of the rim, within about 8-10 hr right ascension, where the cosmic rays intensity exhibits a gradient within a narrow range in right ascension. Although this can be accidental, it is important to note that while it is not possible to decouple a relative modulation in declination (i.e. within right ascension bands) from experimental coverage uncertainties in the experiment local zenith angle, the relative intensity in right ascension (i.e. within declination bands) is experimentally more robust and reliable.

In order to relate the full-sky global anisotropy to the local environment, we schematically show in Fig. \ref{fig:cartoon} the relative location of the excess and deficit regions of cosmic rays with respect to the heliosphere and the LIMF that drapes around it. In the figure the termination shock (TS) is represented as a small circle at the center and the heliopause is represented with a tail toward the left side. The heliotail axis (approximately the direction of the interstellar wind upstream the heliosphere), is in the plane of the figure, and it is coincidentally almost on the ecliptic equatorial plane (which is perpendicular to the figure). In fact the ecliptic coordinates of the heliospheric nose are ($\lambda$, $\beta$) $\approx$ (255$^{\circ}$, 5$^{\circ}$), i.e. just 5 degrees off the ecliptic plane. The LIMF is believed to be approximately directed toward ecliptic coordinates ($\lambda$, $\beta$) $\approx$ (221$^{\circ}$, 39$^{\circ}$) in the upstream direction~\citep{funsten}. This is about 40 degrees from the ecliptic plane and approximately 30 degrees above the plane of the figure, although the uncertainty from different determinations is at least of order 30$^{\circ}$. For simplicity Fig.~\ref{fig:cartoon} schematically shows the component of LIMF in the plane of the figure with its draping around the heliosphere. The full-sky global anisotropy of the 1-10 TeV cosmic rays can be described as composed of a broad relative excess region across the hemisphere south of the magnetic equator (the red shaded sector in the figure, which contains the heliotail and the downstream direction of the LIMF), and a relative deficit region (the blue shaded sector) north of the magnetic equator. The cosmic ray intensity modulation in these regions cannot be described with a dipole, since the transition between the two regions lays approximately along the magnetic equator (the non-colored sectors in the figure) but with the steepest variations in directions with smaller angular distance from the heliotail. 
%As previously described, the magnetic equator seems to be approximately located along the region of steep gradient in the cosmic ray intensity. 
%Observations suggest that above about 100 TeV the anisotropy has a different structure, namely a relative deficit in the region of the sky dominated by a broad excess at lower energies (see~\cite{aglietta,abbasi11}).
At energies in excess of about 100 TeV the heliospheric influence must be sub-dominant, if not inexistent. Whatever the origin of the anisotropy above 100 TeV is, the arrival distribution of cosmic rays at lower energy is likely affected by the heliosphere since the gyro-radius of $\approx$ 8,000 AU is of the order of heliotail size.

\begin{figure}[t]
\begin{center}
\includegraphics[width=\columnwidth]{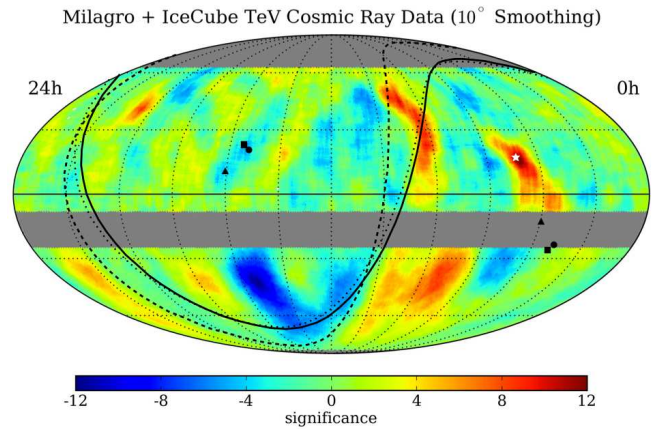}
%"Map in eq coord of the statistical significance ... , where non-zero values (red, blue color) signal non-random fluctuations" or something appropriate to this effect
\caption{\small Map in equatorial coordinates of the statistical significance of the observed cosmic ray counts with respect to that averaged over a right ascension interval $\Delta \alpha$ = 2 - 4 hr. Non-zero values measure how significantly the relative excess and deficit regions (in red and blue, respectively) deviate from random fluctuations (see text). The map shows the observation by Milagro at about 1 TeV in the northern hemisphere~(from~\cite{abdo2}) and that by the IceCube Observatory at a median energy of 20 TeV in the southern hemisphere (from~\cite{abbasi11b}). The symbols are as in Fig.~\ref{fig:LargeScale2}. %Note the different characteristics energies of the cosmic ray particles in the two hemispheres. Also shown the equator "at infinity" (see text) in continuous line, corresponding to the direction of LIMF inferred in~\cite{funsten} (circular symbols), the one in dotted line in~\cite{schwadron} (square symbols), and that in dashed line from~\cite{heerikhuisen} (triangle symbols). The star symbol indicates the downwind interstellar medium flow with coordinate (5 hr, +17$^{\circ}$). 
The map grid is 2 hr in right ascension and 30$^{\circ}$ in declination.}
\label{fig:SmallScale}
\end{center}
\end{figure}

As mentioned above, the global TeV cosmic ray anisotropy appears to have a fine angular structure composed of localized features. Subtracting, from the observation, a map where counts are averaged in each declination band over a right ascension range of $\Delta \alpha$, it is possible to highlight those regions where the cosmic ray intensity has variations on smaller scales than $\Delta \alpha$. In order to assess whether the residual map significantly deviates from random fluctuations of the averaged map, the statistical significance of the observed deviations in relative intensity is calculated using the method described in~\cite{lima83}. In the determination of statistical significance the declination dependence of experimental counts is taken into account, thus providing the most robust determination in declination bands with the largest count (see~\cite{abbasi11b}). Fig.~\ref{fig:SmallScale} shows the map in equatorial coordinates of the statistical significance with respect to the averaged map. The northern hemisphere shows the observation by Milagro at a mean energy of about 1 TeV, where the averaging was done in a 2 hr right ascension range (corresponding to 30$^{\circ}$)~\citep{abdo2}. The southern hemisphere shows the observation reported by IceCube at a median energy of about 20 TeV, with the averaging performed over 4 hr in right ascension (corresponding to 60$^{\circ}$). To obtain the two maps the reference level and residual data counts in each $\sim$1$^{\circ}$ pixel in the sky were added within a radius of 10$^{\circ}$, which therefore represents the smoothing scale (see~\cite{abbasi11b}). Compared to Fig.~\ref{fig:LargeScale2}, this map only shows the small angular scale anisotropy features. As in Fig.~\ref{fig:LargeScale2}, the orientations of the LIMF from~\cite{funsten,schwadron,heerikhuisen} along with their corresponding equator "at infinity" are also shown.

The most significant localized excess region observed in the northern hemisphere (the so-called region A in~\cite{abdo2}) is in the apparent direction of the heliotail, or the downstream interstellar flow direction, with coordinates ($\alpha$, $\delta$) $\approx$ (5 hr, +17$^{\circ}$) (the star in the figure). %In~\cite{reconnection} it is argued that this excess might be related cosmic ray re-acceleration in the stochastic magnetic reconnection regions that accumulate in the heliotail due to the 11-year solar cycle magnetic reversals, with the highest efficiency being along the direction of the heliotail. 
Another significant localized region of multi-TeV cosmic rays observed in the northern hemisphere (the so-called region B in~\cite{abdo2}) appears to be elongated within right ascension $\alpha \approx$ 8 - 10 hr. The excess at about ($\alpha$, $\delta$) $\approx$ (21 hr, +32$^{\circ}$), also visible in Fig.~\ref{fig:LargeScale2} is generated by gamma rays from the Cygnus region, and it is not related to hadronic cosmic rays (see~\cite{abdo2}).
%This edge seems to be within about 10$^{\circ}$ from the equator at infinity of the LIMF as well.
The southern hemisphere shows similar small scale anisotropy features, although with lower statistical significance due to the smaller number of events utilized to compile the map. The most significant and broadest excess and deficit regions observed in the southern hemisphere, at ($\alpha$, $\delta$) $\approx$ (7 hr, -45$^{\circ}$) and ($\alpha$, $\delta$) $\approx$ (15 hr, -45$^{\circ}$), respectively (see Fig.~\ref{fig:SmallScale}) coincide with the center of the corresponding global anisotropy regions in Fig.~\ref{fig:LargeScale2}. This indicates that cosmic ray intensity variation at the relative deficit and excess is sharper than 60$^{\circ}$ scale. As already noted above, the fine structure observed in the north at right ascension $\approx$ 4 - 6 hr does not seem to have a continuity toward the southern sky, which might be a residual effect of the different characteristic energy of the two observations. On the other hand the southern hemisphere elongated feature along right ascension $\alpha \approx$ 8 - 10 hr, is apparently connected to the Milagro region B in the northern hemisphere. We note that such a localized feature does not appear to be particularly significant at this point. However the two elongated localized regions seem to be approximately located at the edge where the broad relative excess of cosmic rays steeply transitions into the deficit region (see Fig~\ref{fig:LargeScale2}). This indicates that the cosmic ray intensity variation across this edge is steeper than 30$^{\circ}$-60$^{\circ}$. In addition a circle drawn over such elongated excess regions would approximately overlap with the relative excess observed in the southern hemisphere at right ascension $\alpha \approx$ 22 - 23 hr, also located by the region of largest intensity gradient, although it is not statistically significant and it might be a coincidence. In any case, the elongated localized fractional excess region across a big portion of the sky, is approximately parallel and close to sightlines perpendicular to the LIMF direction inferred from several indirect determinations (see Figs.~\ref{fig:LargeScale2} and~\ref{fig:SmallScale}). %Such an arc-like structure of the TeV cosmic rays is reminiscent of the ribbon of keV ENAs mapped by IBEX~\citep{funsten,schwadron,heerikhuisen}, and of which it approximately shares the circular center.

%The more ecliptic southward location of the TeV arc might indicate that its formation is farther away from the heliosphere, where scattering processes become important.

% POSSIBLE ARTIFACTS FROM EXPERIMENTAL DETERMINATION OF THE MAPS
It is important to note that the anisotropy observations at small angular scales might be affected by possible artifacts arising from the experimental techniques used. On the other hand, based on the current results we find the correlation of the region of steepest cosmic ray intensity variations with the interstellar magnetic equator in the 1-10 TeV energy range, compelling, and as a consequence we could relate some of the observed localized fractional excess regions, at least, to processes occurring within or in the outer boundary of the heliosphere.

%-----------------------------------------------------------------------------------------------------------------------------------------------
\section{Probing the outer Heliosphere and the LIMF}
\label{sec:probe}

%In diffusive regime, where propagation along distances sufficiently longer than the mean free path are taken into account, cosmic ray propagation perpendicular to the mean magnetic field is possible due to the field turbulent wandering at large scale. 
Cosmic rays transport in the interstellar magnetic field is diffusive as long as they propagate through distances larger than the scattering mean free path, beyond which particles lose memory of their initial direction, and can propagate across magnetic field lines due to their stochastic wandering at a scale of 10-100 pc. In the energy range of 1-100 TeV cosmic rays have a scattering mean free path in the interstellar medium of order 1-10 pc~\citep{yan08}, therefore at smaller distances the diffusion regime is broken as particles velocities are not de-correlated. This means that particles mostly stream along the LIMF within the interstellar mean free path. We assume that below about 100 TeV cosmic ray arrival distribution is similar to that at higher energy. Whatever causes the anisotropy observed at several 100's TeV, there is no reason to believe that at lower energy it has disappeared or significantly changed by purely interstellar propagation. Therefore, at energies where the heliospheric influence becomes important, as anisotropic cosmic rays approach the heliosphere, they begin to feel the turbulent perturbations $\delta B$ induced by the wake, with exponentially increasing amplitude the smaller the distance. At the largest scales, turbulence is super-Alfv\'enic, therefore it is strong ($\delta B/B_0 \approx 1$, with $B_0\sim 3\mu$G the regular magnetic field) and isotropic (see Sec.~\ref{sec:helio}). Particles propagating through the turbulent magnetic field are affected by scattering processes that change their pitch angle, with a rate that depends on $(\delta B/B_0)^2$. Therefore, scattering rate increases fast the closer to the heliosphere, leading to a progressively more uniform pitch angle distribution.
%Since pitch angle scattering frequency in isotropic turbulence goes as $D_{\mu\mu} \propto (\delta B / B_0)^2$~\citep{tautz06,tautz08}, it exponentially increases as cosmic rays approach the outer heliosphere. 

%Cosmic rays predominantly scatter on slowly varying turbulent field modes that have the component of wave vector parallel to the mean magnetic field such that $k_{\parallel} v_{\parallel} \approx \Omega$, with $v_{\parallel}$ the parallel particle velocity $\vec{v}$ and $\Omega \approx 1.5\times 10^{-3}Z\,(B_0 / \mu G)$ Hz its gyro-frequency.
Cosmic rays predominantly scatter on magnetic field fluctuations that have a size scale $L \approx r_g$, with 
\begin{equation}
r_g = \frac{p_{\perp}}{ZeB} \approx \frac{200}{Z}\,\left(\frac{E}{1\, TeV}\right)\,\left(\frac{1\, \mu G}{B}\right) AU
\label{eq:energy}
\end{equation}
the particle gyro-radius, where $p_{\perp}$ is the momentum perpendicular to the magnetic field direction, $E$ the particle energy, and $B = B_0+\delta B$ the total magnetic field of the perturbation. %Magnetic fluctuations at the largest scales $L\approx 100-600$ AU carry most of the turbulence power and therefore in general they have the dominant contribution to resonant scattering processes. 
Considering that $\delta B/B_0 \approx 1$, this resonance condition is satisfied for proton energies of approximately 10 TeV at the largest scales (i.e. $\approx$ 300-600 AU). This energy scale effectively determines the separation between two regimes, as discussed in the following subsections. %At lower energy the particles with smaller gyro-radius have a reduced pitch angle scattering with the large scale perturbations, and resonant scattering with the cascading turbulence is the most important process. At higher energy the magnetic perturbations do not affect particle trajectory as much any more.
In resonant scattering with isotropic turbulence, particles pitch angle changes randomly by an amount of order $\delta B/B_0$ in each collision~\footnote{which is the average inclination of the field lines from the mean field direction due to the magnetic irregularities.}. Therefore, when the average effective deviation in pitch angle after $N$ collisions is large (i.e. particles have scattered approximately 90$^{\circ}$ from the initial direction, or equivalently $\sqrt{N}\, (\delta B/B_0)\approx 1$), particles have lost memory of their initial condition. The effective mean free path, defined as the distance at which pitch angle randomization is realized, is then
\begin{equation}
\lambda \approx N\,L \approx r_g\,\left(\frac{B_0}{\delta B}\right)^2.
\label{eq:lambda}
\end{equation}
Therefore, resonant scattering in strong turbulence at the largest scales generates a re-organization of the cosmic ray arrival direction distribution. %, that may depend on the initial direction and amplitude of the anisotropy. %independently of their initial distribution. %Contributions from resonant scattering of lower energy particles on the cascading turbulence at smaller scales is also important, although with a reduced rate corresponding to the smaller magnetic fluctuations $\delta B$.
However, due to the super-Alfv\'enic turbulence distribution in the outer heliosphere at the largest scales, and the relative angle between the LIMF and the heliotail, the effective scattering rate depends on the arrival direction of the cosmic rays, and pitch angle scrambling is more important toward directions with the largest scattering. This may explain the complex angular structure of the observed TeV anisotropy, which could still hold some information on the direction and amplitude of the initial underlying anisotropic distribution.
%no matter what the initial anisotropic distribution of the cosmic rays is.
%On the other hand the observed anisotropy at energy in excess of about 100 TeV might hold information on the LIMF on a larger distance scale than the heliospheric size.

\subsection{Scattering at about 10 TeV}

\begin{figure}[t]
\begin{center}
\includegraphics[width=0.9\columnwidth]{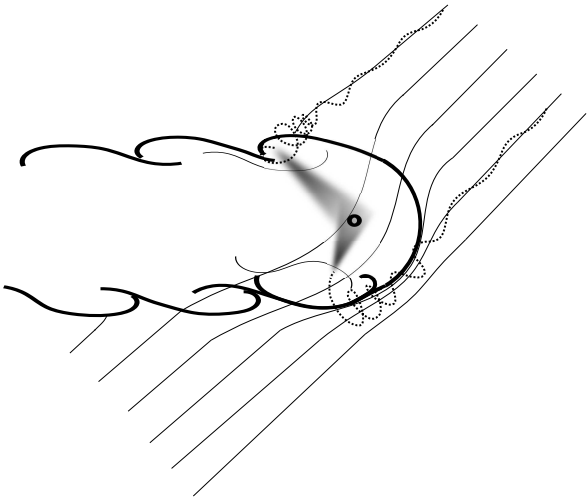}
\caption{\small Illustration of the turbulent ripples at large scale on the outer heliospheric boundary. High energy cosmic rays propagating from the upstream direction into the front of the heliosphere resonantly scatter with the trailing magnetic field perturbations on the flanks of the heliosphere and along the heliotail. The uniform pitch angle distribution makes it possible for particles to scatter back to Earth. The grey blurred area schematically represents the possible scattered particle trajectories toward Earth.}
\label{fig:scattering}
\end{center}
\end{figure}

Protons at $\approx$ 10 TeV have resonant scattering with the isotropic super-Alfv\'enic turbulence at the largest scales. Particles streaming along the LIMF from the downstream direction (i.e. from the lower left side in Fig.~\ref{fig:cartoon}) are expected to experience higher scattering rate while they get closer, due to the angular proximity to the heliotail. Because of the increasingly strong magnetic field perturbations, resonant scattering scrambles high energy cosmic rays with the highest rate occurring at smaller impact distance from the heliotail. Cosmic rays that pass into the upstream region have very little chance to back-scatter, because the super-Alfv\'enic turbulent perturbations cannot propagate in that direction. In particular at the largest scales, turbulence trails downstream at a mean angle of approximately 40$^{\circ}$ from the interstellar flow direction. %, which is coincidentally close to the inferred angle of the LIMF direction.
%This causes an effective excess of cosmic rays from the downstream direction with a gradient in intensity along sightlines where resonant scattering becomes inefficient (i.e. approximately at directions perpendicular to the LIMF in the outer heliosphere). 
For the same reason, energetic particles streaming along the LIMF from upstream, experience very low scattering rate in the outer heliosphere until they reach the downstream region, where they can resonantly scatter and scramble their distribution, feeding and enhancing the downstream flux. This is illustrated in Fig.~\ref{fig:scattering}, where two energetic particles streaming along the LIMF from the nose side of the heliotail, and affected by the turbulent magnetic field, scatter at the flanks of the heliosphere and along the heliotail back to Earth. The uniform pitch angle scattering distribution scrambles particle directions and the possible trajectories toward Earth are represented by the blurred grey areas. If the turbulence surrounding the flanks of the outer heliosphere and the heliotail extends into the LISM for about 100 AU, resonant scattering can generate a flux of back-scattered particles that may be responsible for the re-distribution of the initial anisotropy, and in particular for the excess toward the heliotail. A quantitative estimate of back-scattered particles compared to the direct flux is done in Sec.~\ref{ssec:simplemodel} using a simplified model. This model shows that the overall effect of scattering on the turbulent outer heliosphere can be that of generating the observed asymmetry in arrival distribution. Because of the angle between the LIMF and the heliotail and of the trailing of turbulence at the largest scales, the gradient in the cosmic ray intensity is approximately located toward sightlines that are perpendicular to the mean LIMF. Such gradient is expected to be larger where the LIMF hit the heliosphere at a steep angle (i.e. the upper left side in Fig.~\ref{fig:cartoon} and~\ref{fig:scattering}, corresponding to right ascension range of $\alpha \approx$ 8-10 hr in Fig.~\ref{fig:LargeScale2}), while the intensity has a smoother variation where the LIMF tangentially drapes the heliosphere (for instance at the lower right side in the figures, at $\alpha \approx$ 18-22 hr) because only smaller and weaker turbulent perturbations can effectively scatter cosmic rays.
%***
%No matter what the initial arrival distribution of the cosmic rays, such resonant scattering processes on the super-Alfv\'enic isotropic turbulence in the outer heliosphere can be significantly strong, and an initial cosmic ray anisotropy of $\sim$10$^{-3}$ is re-distributed differently with similar amplitude.

%The new distribution is asymmetric with a broad relative excess in the downstream interstellar flow, effectively delimited by perpendicular sightlines to the LIMF, and with the highest intensity toward view diorections closer to the heliotail.

The line-of-sights normal to the mean LIMF, can be several degrees away from the magnetic equator "at infinity", because of the LIMF deformation due to the heliosphere. In fact, the locus of largest gradient in the cosmic ray arrival direction distribution, appears to be within about 30$^{\circ}$ from the magnetic equator inferred from indirect estimations of the LIMF direction. In summary, the relative angle between the LIMF and the heliotail, although still uncertain, and the trailing turbulent magnetic perturbations at the largest scale, are responsible for the non-dipolar nature of the global anisotropy.

As mentioned in Sec.~\ref{sec:obs}, a steep intensity variation in the cosmic ray arrival direction distribution, may appear like localized fractional excess regions when a residual map is derived. The width of such localized regions is a function of the magnitude of the gradient, and it depends on the averaging procedure of the cosmic ray map. However, in this model, it is the existence of such an intensity gradient that has a physical relation to the scattering processes of cosmic rays with the super-Alfv\'enic magnetic field turbulence on the outer heliospheric boundary. The observation could have some relation with the ribbon of keV ENAs observed by IBEX, attributed to some mechanism that enhances ENA flux in the sight of lines approximately perpendicular to the LIMF as it drapes around the heliosphere. In particular,~\cite{schwadron12} considered cosmic ray particles diffusion through the magnetic field draping around the heliosphere as the cause of the observed anisotropy, within a consistent model that addresses the IBEX ribbon. The more ecliptic southward location of the TeV cosmic ray observation might indicate that its formation is farther away from the heliosphere, where scattering processes become important.

\subsubsection{Effect of particle back-scattering}
%\label{app:simplemodel}
\label{ssec:simplemodel}

\begin{figure}[t]
\begin{center}
\includegraphics[width=1.0\columnwidth]{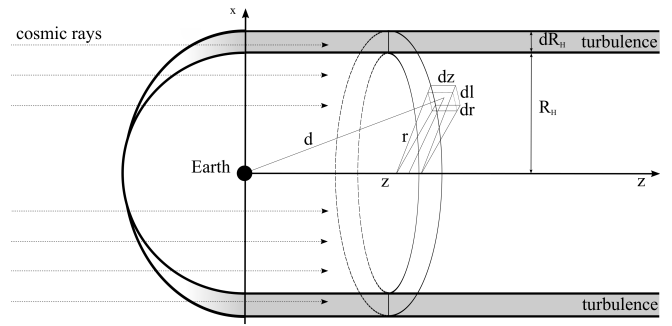}
\caption{\small Schematic model of cosmic ray scattering within a shell of turbulent plasma in the outer heliosphere. Energetic particles with guiding center streaming along the interstellar flow direction scatter inside the turbulent shell back to Earth. Due to resonant scattering and the length of the heliotail, back-scattered particles can be of the same order of magnitude of the direct flux, or even slightly in excess.}
\label{fig:simplemodel}
\end{center}
\end{figure}

Here, a simple model of the effect of cosmic ray back-scattering on the turbulent outer heliosphere along the tail is presented. For simplicity we assume that the guiding centers of the cosmic rays stream along the interstellar flow (i.e. along the z-axis in Fig.~\ref{fig:simplemodel}, with its origin corresponding to the location of Earth). In this model we assume that resonant scattering happens only within a shell around the heliotail of thickness $dR_H$ (the grey shaded area in the figure), where the plasma is fully turbulent. The flux of cosmic rays can be separated in an isotropic and an anisotropic contribution. Since the isotropic component remains as such, it is the anisotropic part of the flux that gets re-distributed by scattering processes.

The anisotropic component of the cosmic rays, with the relative excess approximately toward the direction of the heliospheric nose (i.e. in the up-stream interstellar flow direction), is represented with flux density $n_{CR}$. When these particles pass through the turbulent heliospheric shell, they can have resonant scattering (depending on their energy and the turbulence scale). In particular, particles that reach an infinitesimal volume $dV=dr\,dl\,dz$ of turbulent plasma past the Earth along the tail, are uniformly re-distributed and are back-scattered to Earth.

The number of cosmic ray particles inside $dV$ at a given time is

\begin{equation}
dN=n_{CR}\,dr\,dl\,dz.
\label{eq:dn}
\end{equation}
If $P_s$ is the scattering probability, the number of particles uniformly scattered from the infinitesimal volume $dV$ is

\begin{equation}
dN_s=P_s\,dN.
\label{eq:dns}
\end{equation}
Therefore, the number of particles that reach Earth from $dV$ is

\begin{equation}
dN_b=dN_s\,\frac{R^2_E}{d^2},
\label{eq:dne}
\end{equation}
where $R_E\sim 4.3\times 10^{-5}$ AU is the Earth's radius, and $d=\sqrt{z^2+r^2}$ is the distance of $dV$ from Earth. Combining Eqs.~\ref{eq:dn},~\ref{eq:dns},and~\ref{eq:dne} we obtain

\begin{equation}
dN_b=n_{CR}\,P_s\,R^2_E\,\frac{dr\,dl\,dz}{z^2+r^2}.
\label{eq:dne2}
\end{equation}

Since we want to calculate the amount of cosmic rays that scatter back to Earth, we integrate Eq.~\ref{eq:dne2} over the volume of the turbulence shell of radius $R_H$ and thickness $dR_H$ and along positive values of z (i.e. past the Earth, see Fig.~\ref{fig:simplemodel}). So the number of cosmic rays back-scattered to Earth is

\begin{eqnarray}
N_b & = & n_{CR}\,P_s\,R^2_E\,\int_{R_H}^{R_H+dR_H}dr\,\int_0^{2\pi r}dl\,\int_0^{\infty}\frac{dz}{z^2+r^2}
\nonumber \\ \nonumber \\
        & = & n_{CR}\,P_s\,\pi^2\,R^2_E\,dR_H,
\label{eq:dnefinal}
\end{eqnarray}
where the heliotail is long enough (about 10$^3$-10$^4$ AU) that we can approximate the integration on z to infinity. Eq.~\ref{eq:dnefinal} implicitly assumes that the contribution of cosmic rays that back-scatter from the far regions of the heliotail is taken into account. From the integral in z in Eq.~\ref{eq:dnefinal} we see that about 50\% of particles back-scattered to Earth comes from within a distance of about 300 AU along the heliotail. Cosmic rays reaching that distance and propagating back to Earth take about $\tau \sim 6\times 10^5$ sec, which is considered as the back-scattering time scale. In this time the number of cosmic rays directly reaching Earth (i.e. from the negative values of z) is

\begin{equation}
N_d=n_{CR}\,4\pi R^2_E\,c\,\tau.
\label{eq:nd}
\end{equation}

The asymmetry derived from the back-scattered particles with respect to the direct flux can be evaluated with

\begin{equation}
\delta = \frac{N_b-N_d}{N_b+N_d}=\frac{N_b/N_d-1}{N_b/N_d+1},
%\frac{\frac{3\pi P_s dR_H}{4 c\,\tau}-1}{\frac{3\pi P_s dR_H}{4 c\,\tau}+1}.
\label{eq:delta}
\end{equation}
where

\begin{equation}
\frac{N_b}{N_d}=\frac{3\pi}{4}\,P_s\,\frac{dR_H}{c\,\tau}
\end{equation}

Based on this model, the back-scattered cosmic rays exceeds the direct flux (i.e. $\delta \gtrsim 0$, meaning that the initial anisotropy is completely re-directed approximately toward the heliotail) if $P_s \gtrsim 100/dR_H$. Therefore if we assume that $dR_H \sim 100$ AU, the scattering probability must be very close to 100\%. Although it is reasonable to assume that at the resonance $P_s \sim 100\%$, experimental observations are affected by energy resolutions that are not negligible, therefore they cover ranges where the effects of resonant scattering are reduced, thus decreasing the effective scattering probability. Therefore, based on this model, in order to have enough back-scattered particle to re-organize the cosmic ray arrival distribution at a given energy, the turbulent shell should be wider than 100 AU. However, the energy resolution of the observations relaxes such condition by allowing particles at different energies to contribute to the back-scattered flux.

%Cosmic rays propagating from the direction of the heliotail through the turbulent shell contribute to increase the flux from the back, relaxing the constrains on the range of $dR_H$ and $P_s$.

%If we assume that $dR_H\sim 100$ AU (see Sec.~\ref{sec:helio}), then the back-scattered cosmic ray flux exceeds the direct flux by $\delta \sim 10^{-4}-10^{-2}$, for scattering probability $P_s \sim 1-100\,\%$, respectively. At the resonance scattering fully scrambles particle directions, therefore $P_s = 100\,\%$. On the other hand, experimental observations are affected by energy resolutions that are not negligible, therefore they cover ranges where the effects of resonant scattering are attenuated, thus reducing the effective scattering probability. Therefore, in the TeV energy range, resonant scattering at the largest scale, is reduced due to the smearing of the energy determination. While at sub-TeV and over $\sim$ 100 TeV energy range, resonant scattering is sub-dominant and the asymmetry effect is reduced.
We believe that the simple assumptions made in this model, do not affect the general evaluation of the effects of cosmic ray resonant scattering with the turbulent outer heliosphere along the tail. On the other hand, the complexity of the heliospheric structure and of the turbulence on its outer boundary, and the fact that cosmic rays mainly stream along the interstellar magnetic field, which is at an angle with respect to the heliotail, requires full numerical simulations in order to provide a solid prediction of the effects that scattering processes have on TeV cosmic rays.

\subsection{Scattering below 10 TeV}

Cosmic rays in the 1-10 TeV energy range can have resonant scattering with the cascading isotropic turbulence at scales down to order of 100 AU. The decrease in power of the smaller turbulence scales is compensated by the higher cosmic ray density at lower energy, so that effectively resonant scattering is still significant, and the anisotropy has the same topology and similar amplitude.

Sub-TeV cosmic rays can have resonant scattering only at scales where turbulence is sub-Alfv\'enic. Since at such small scales cascading turbulence does not have time to develop a significant anisotropy, resonant scattering at scales of order 10 AU is a possible important contribution. However damping processes might be large enough to limit turbulence inertial range at this scale. On the other hand an important contribution comes from particle propagation through larger turbulence scales. In this case particles trajectory follow their guiding center along magnetic field lines and the pitch angle varies gradually with the changes in magnetic field due to the conservation of the first adiabatic invariant $p_{\perp}^2/B$. An increase of B over times longer than particle's gyration time, produces an increase in the transversal momentum, leading to an effective change in pitch angle. In this case pitch angle scattering is not completely uniform, leading to a similar anisotropy topology, but with reduced amplitude, as confirmed by observations (see for instance~\cite{amenomori05}).

The mechanism where relatively low energy cosmic rays are affected by larger turbulence scales, where non-resonant scattering can be important in scrambling cosmic ray arrival directions (see also~\cite{schli98}), was considered by~\cite{yan04,yan08,yan11,beres11}. In particular, such non-resonant processes with large turbulence scale produce gyro-resonance instabilities that effectively transfer turbulence energy into additional small scale Alfv\'enic waves, thus increasing low energy cosmic ray scattering rate~\citep{LB06,yan11}. This mechanism might not play the same role in the case of small inertial range, although this would require a more dedicated study with numerical simulations. In any case, the fact that anisotropy observations below 1 TeV provide evidence of a tail-in excess, supports the idea that scattering processes are likely important in the sub-TeV energy range as well.

\subsection{Scattering above 10 TeV}

At energies above about 10 TeV, proton gyro-radius is greater than the largest turbulence scale. In this case particles move in orbits determined by the mean magnetic field, and the resonant scattering with magnetic fluctuations starts to be sub-dominant. Lower scrambling of cosmic rays pitch angle means a decreasing anisotropy, as observed. In particular, the intensity gradient by the LIMF equator becomes progressively more gradual and, eventually, the small scale features broader and indistinguishable from the global anisotropy distribution. Non-resonant scattering with turbulence at scales much smaller than the gyro-radius can have important effects as well, independently of the degree of turbulence anisotropy, although this case has not been studied in detail yet.

For energies higher than about 100 TeV, even the proton gyro-radius is larger than the heliotail and cosmic ray anisotropy is dominated by interstellar effects within distance scales of their mean free path in the interstellar medium.

Heavier cosmic rays can contribute to resonant scattering processes at higher energies, possibly producing significant effects from heliospheric perturbations well above 100 TeV, depending on mass composition. For instance cosmic rays are expected to be dominated by helium nuclei above about 10 TeV~\citep{cream,pamela,argo12}, and to have significant contribution from heavier particles at $\approx$ PeV range (i.e. the knee region)~\citep{hoerandel03}. The evolution to heavier mass composition, inevitably has the effect to smear the energy transition between heliospheric and interstellar anisotropy by a few hundreds TeV (see~\cite{aglietta,abbasi11}).

Observations of cosmic ray anisotropy with experiments that have a wide energy response, result in averaging scattering contributions at different scales, and therefore in smearing the resonant scattering effect. The resulting sky maps inevitably show anisotropies originated at different energies and make it difficult to disentangle the individual effects. However, current observations make it still possible to provide good order of magnitude comparisons with models.

\subsection{Scattering in the inner heliotail}

Cosmic rays in the GeV to TeV energy range, propagating through the inner heliotail, are also affected by the complex magnetic field structure, with $\sim$200 AU unipolar domains generated by solar cycles, and strongly mixed magnetic field regions in between. Stochastic magnetic reconnection processes may cause re-acceleration as long as particles can be trapped inside the reconnecting domains, and scattering processes are not sufficiently strong to cause significant energy re-distribution (see~\cite{reconnection,recon2}). Such scattering processes are weaker for those energetic particles propagating at low solar latitudes, where the magnetic field is strongly mixed at small scale (see Sec.~\ref{sec:helio}). Away from this directions scattering with the larger unipolar domains scramble TeV particles in a broader distribution, providing an environment where acceleration by magnetic reconnection is less effective. This may be at the origin of the localized fractional excess region observed in the interstellar flow downstream direction and of the claimed spectral anomaly~\citep{abdo2,argo12}.

Observations of a correlation of low energy cosmic rays with the solar cycles, have been documented and understood as due to the influence of the variable magnetic field intensity and turbulence from solar activity. An anisotropic effect correlated with solar cycles was observed with 600 GeV cosmic rays as well, and interpreted in terms of acceleration due to the solar wind induced electric field expected when cosmic rays cross the solar magnetic field neutral sheets~\citep{munakata11}. Recently, a $\sim$ 4$\times$10$^{-3}$ relative modulation of TeV muon intensity (produced by $\approx$ 10 TeV cosmic rays) correlated with the solar cycle, superimposed to the well known seasonal variation due to the upper atmosphere's temperature, was reported based on over 20 years of muon measurements at the Gran Sasso Laboratory~\citep{muons}. If this observations will be confirmed by other long term measurements of multi-TeV cosmic rays, it might provide a confirmation that at such energies particles from the direction of the heliotail may actually be influenced by the solar magnetic field polarity domains within a distance of about 300 AU.

%-------------------------------------------------------------------------------

\section{Conclusions}
\label{sec:conclusion}

TeV cosmic ray anisotropy can be the result of accidental galactic sources that happened to accelerate particles at relatively close distance in relatively recent times. This intriguing scenario that relates the observed anisotropy to the origin of galactic cosmic rays can be used to provide constraints on particle acceleration at the source and propagation across the galactic disk, if diffusive regime is assumed. %~\citep{blasi}.
These models should be able to consistently describe the diffuse $\gamma$ ray emissions from the interactions of cosmic rays with the interstellar medium, molecular gas, dust and radiation and magnetic fields, as well. On the other hand the uncertainties on the source distribution, and the simple assumption of single power law scattering frequency, and on their dependency on the type of galactic medium, make such predictions susceptible of large uncertainties. The non-dipolar structure of the global cosmic ray anisotropy and its fine angular structure challenges theories relating this observations with large scale diffusive models. Even though it is likely that one or more nearby sources imprint an anisotropy, non diffusive cosmic ray propagation at distances smaller than their mean free path can easily re-distribute their arrival direction. Within such a propagation volume, stochastic fluctuations in the interstellar medium magnetic field can generate random fluctuations in the cosmic ray anisotropy that appear as localized excess regions.%~\citep{giacinti}.

While awaiting for more refined experimental observations of cosmic ray anisotropy as a function of energy and angular structure from various experiments and with higher event statistics, it seems natural to assume that the elongated heliosphere must affect in some way cosmic rays with gyro-radius comparable to its size. Cascading turbulence from the wake, which trails downstream the interstellar flow and outward into the LISM, resonantly scatter TeV cosmic rays streaming along the LIMF. The angle between the heliotail and the LIMF and the asymmetry in the turbulence from the up-stream and downstream direction produces the non-dipolar anisotropy observed below about 100 TeV. At higher energy the anisotropy has a different structure and is most probably affected by propagation processes at larger scales, or even have an imprint from one or more galactic sources. The structure of TeV cosmic rays is related to the interplay between the LIMF and the heliosphere, and therefore it can be used to infer information on the interstellar magnetic field direction just outside the heliospheric boundary (i.e. within several hundreds AU). The small scale features that arise where the gradient of cosmic ray intensity is sufficiently large within 30$^{\circ}$-60$^{\circ}$ angular range, highlight a possible connection to scattering with the turbulent magnetic field.

In order to validate such model, dedicated numerical MHD simulation models of the interaction between the interstellar and heliospheric magnetic field, and of the their turbulence, need to be developed. Current MHD simulations of the heliosphere have reached rather high complexity (see for instance~\cite{pogorelov,pogorelovb}).

Although in general it is not trivial to combine observations from various experiments to complete the sky coverage, because of the different energy response and analysis techniques, it is possible now to have a first view into the global anisotropy and its small angular scale structure in the 1-10 TeV range. Dedicated observations with comparable energies from different parts of the sky can provide more solid experimental evidence to test the models. In the meantime we consider the model presented here as a simple explanation of the observations, that will need verification as progresses in heliospheric modeling and more experimental results are available. The simple quantitative model presented in Sec.~\ref{ssec:simplemodel} is just able to provide a glance into the proposed mechanism. Due to the complexity of the heliospheric structure and the turbulence that it entails, only numerical simulations can provide a solid prediction of the effects that scattering processes have on TeV cosmic rays.

\acknowledgments

The authors would like to thank the anonymous reviewer for helping improve the paper. Many thanks to Priscilla Frisch for the useful discussions on the impact that an observation in TeV cosmic rays might have in the understanding of the properties of the outer heliosphere. The authors wish to thank the colleagues at WIPAC and the Department of Astronomy for discussions on cosmic ray anisotropy, and Nikolai Pogorelov for useful discussions on MHD simulations of the heliosphere. PD acknowledges the support from the U.S. National Science Foundation-Office of Polar Programs. AL acknowledges the support of the NSF grant AST-1212096, NASA grant X5166204101 and of the NSF-sponsored Center for Magnetic Self-Organization.

%-------------------------------------------------------------------------------

%\bibliographystyle{unsrt}
\bibliographystyle{apj}
\bibliography{papers}

\end{document}